\documentclass{article}

\PassOptionsToPackage{numbers, compress}{natbib}

\usepackage[preprint]{neurips_2026_vericode}

\usepackage[utf8]{inputenc}
\usepackage[T1]{fontenc}
\usepackage{url}
\usepackage{booktabs}
\usepackage{amsmath}
\usepackage{amsfonts}
\usepackage{nicefrac}
\usepackage{microtype}
\usepackage{xcolor}
\usepackage{graphicx}
\usepackage{float}
\usepackage{needspace}
\usepackage{placeins}
\usepackage[hidelinks]{hyperref}
\hypersetup{
  pdftitle={Evaluating Exact Output and Checkpoint-State Prediction in Real Programs},
  pdfauthor={Xiaohong Chen, David Bucur, Chenglong Ma, Yi Zhang, Lingming Zhang, Sriram Vishwanath, Grigore Rosu}
}
\graphicspath{{figures/}}

\newcommand{\Occ}{\ensuremath{\Omega_{\mathrm{CC}}}}
\newcommand{\Otrace}{\ensuremath{\hat{\Omega}_{\mathrm{Trace}}}}
\newcommand{\Osize}{\ensuremath{\hat{\Omega}_{\mathrm{StateSize}}}}
\newcommand{\Oload}{\ensuremath{\hat{\Omega}_{\mathrm{StateLoad}}}}
\newcommand{\lcbpy}{\texttt{lcb\_python}}
\newcommand{\lcbcpp}{\texttt{lcb\_cpp}}
\newcommand{\classic}{\texttt{classic}}
\newcommand{\cc}{\texttt{codecontests}}
\newcommand{\nx}{\texttt{networkx}}

\title{Evaluating Exact Output and Checkpoint-State Prediction in Real Programs}

\author{%
  Xiaohong Chen\textsuperscript{1} \quad David Bucur\textsuperscript{1}
  \quad Chenglong Ma\textsuperscript{1} \quad Yi Zhang\textsuperscript{1}\\
  \textbf{Lingming Zhang}\textsuperscript{2}\thanks{These authors took part
    in this work in collaboration with Intent Computing, Inc., outside
    their university duties and without university funds, facilities,
    students or contracts.}
  \quad \textbf{Sriram Vishwanath}\textsuperscript{1,3}\footnotemark[1]
  \quad \textbf{Grigore Rosu}\textsuperscript{1,2}\footnotemark[1]\\
  \textsuperscript{1}Intent Computing, Inc.\\
  \textsuperscript{2}University of Illinois Urbana-Champaign
  \quad \textsuperscript{3}Georgia Institute of Technology\\
  \texttt{\{xiaohong,david,chenglong,yi\}@intentcomputing.org}\\
  \texttt{\{lingming,grosu\}@illinois.edu}
  \quad \texttt{sriram@ece.gatech.edu}\\
}

\newcommand{\FundingDisclosure}{%
  This work received no funding. The authors declare no competing financial
  interests beyond their listed primary affiliations.
}

\begin{document}

\maketitle

\begin{abstract}
We present a benchmark for predicting final output and checkpoint state from source and input alone. It extends CRUXEval-style output prediction with paired shorter- and longer-trace inputs and checkpoints inside and after a loop. The benchmark contains 400 cases from 371 Python and C++ programs, evaluated under seven settings from four model families without tools or code execution. Of 11,200 planned predictions, 11,151 produced gradable responses. Reasoning-enabled settings outperform their off counterparts by 33.1 to 55.2 percentage points on completed responses. The strongest setting scores 93.0\% on shorter-trace final output, 77.0\% on longer-trace final output, and 65.5\% and 63.5\% on the two state tasks; these scores also hold when missing responses count as wrong. Across 2,397 matched Python comparisons with identical source, changing to the longer-trace input yields 528 correct-to-wrong changes and 147 reversals. Source-clustered analyses preserve this accuracy gap, while adjusted Python models give no evidence of a positive incremental association between cumulative state load and error. Changed inputs and checkpoint tasks alter several factors together, so the gaps do not isolate trace length or an internal state-tracking mechanism. The benchmark exposes errors hidden by short-output scores alone.
\end{abstract}

\section{Introduction}

Coding agents combine a language model with a system that supplies context, runs code, exposes feedback, and organizes retries. SWE-agent, OpenHands, and Agentless show how these choices shape performance on software-engineering tasks~\citep{sweagent,openhands,agentless}. Their end-to-end scores measure the resulting model--harness combination. A complementary capability is the model's ability to determine program behavior directly from source and input. Measuring this capability helps identify which execution questions a model can answer before runtime feedback enters the workflow.

Direct execution reasoning remains less extensively benchmarked than code generation. A recent review of 142 papers identifies 156 code benchmarks across 32 tasks, with 86 devoted to code generation~\citep{benchmarksurvey}. CRUXEval-X identifies the same imbalance~\citep{cruxevalx}. We focus on \emph{model-only} evaluation, in which a served endpoint receives source and input with tools, retrieval, and code execution disabled. This interface isolates the prediction task while retaining the endpoint's requested reasoning mode.

CRUXEval gives this question a clean form~\citep{cruxeval}. Its 800 generated Python functions are 3--13 lines long, and each supports output prediction from an input and input prediction from an output. The benchmark separates code execution reasoning from program synthesis. Meta's Code World Model (CWM) approaches the same capability from training~\citep{cwm}. CWM is mid-trained on Python interpreter traces and trajectories from agentic execution environments, then evaluated on trace prediction and coding tasks. In the reported 8B ablation, adding Python traces raises CRUXEval output prediction from 44.6\% to 73.9\% after mid-training, while SWE-bench Verified changes from 18.6\% to 18.4\% after subsequent supervised fine-tuning. The different responses to trace supervision support measuring execution prediction separately from end-to-end agent performance.

We follow CRUXEval's output-prediction task and extend it to longer executions and intermediate state. Each case produces four related questions. Two show byte-identical clean source and request final output under inputs selected to induce shorter and longer native traces. Two use the longer input and instrumented source, stopping at a declared checkpoint inside a loop or immediately after the loop exits. Figure~\ref{fig:running-example} shows an admitted case from the Python LiveCodeBench study.

\begin{figure}[H]
\centering
\begin{minipage}{0.96\linewidth}
\small
\textbf{Clean core loop}\quad\emph{input parsing omitted}
\begin{verbatim}
cnt0 = cnt1 = 0
ans = 0
for ch in S:
    if ch == '0':
        dp0 = 1
        dp1 = cnt0 + cnt1
    else:
        dp0 = cnt1
        dp1 = cnt0 + 1
    ans += dp1
    cnt0, cnt1 = (dp0, dp1)
print(ans)
\end{verbatim}
\footnotesize
\begin{verbatim}
short-trace-final   N=5, S=00110       -> 9
long-trace-final    N=1002, S=(01)^501 -> 251502
inside-loop-state   before iteration 502
                    -> {"ans":63250,"cnt1":500,"dp1":500}
post-loop-state     after the loop
                    -> {"ans":251502,"cnt1":2,"dp1":2}
\end{verbatim}
\end{minipage}
\caption{The four benchmark arms for the real \texttt{abc310\_e} case. The notation $(01)^{501}$ abbreviates the literal 1,002-character input. The inside checkpoint fires after 501 iterations have completed. Both state arms emit canonical JSON from instrumented copies of the clean source.}
\label{fig:running-example}
\end{figure}

The recurrence tracks how many substrings ending at the current position have iterative NAND value zero or one, and \texttt{ans} accumulates the one-valued count. The short-input arm in Figure~\ref{fig:running-example} requires five loop iterations, while the long-input arm applies the same recurrence 1,002 times. The state arms ask for the same three fields midway through and after that execution. These related questions separate success on a short final-output task from success on a longer execution, including the prediction of its intermediate values.

We construct 400 such cases from 371 distinct Python and C++ programs, spanning contest solutions, textbook algorithms, and a library routine. Execution checks establish a stable ordering between the two trace lengths, reachable and deterministic checkpoints, and preservation of the clean program's normal output under instrumentation. Seven model settings answer the resulting tasks in fresh sessions. The benchmark provides paired execution scales and shared state projections per case. We report the collection and grading protocol, aggregate results, source-clustered analyses, and a worked example in this paper.

The central result is that reasoning substantially improves execution prediction, yet longer-input and checkpoint tasks remain difficult. Reasoning-enabled settings exceed their off counterparts by 33.1 to 55.2 percentage points overall. Even the strongest setting falls from 93.0\% on short final-output tasks to 77.0\% on long final-output tasks and below 66\% on both state tasks. In 2,397 paired Python comparisons, the longer input changes 528 correct answers to wrong ones and 147 wrong answers to correct ones, with source held byte-identical.

We then examine how errors relate to static source structure and dynamic measurements. Cumulative state load \(L=\sum_t |S_t|\) combines observed steps with reachable state at each step. Original pooled associations weaken within individual task conditions. We add analyses that account for shared source programs and adjust for task, study, model setting, trace length, and peak state size. These provide no evidence of a positive incremental state-load association. Cyclomatic complexity has no consistent positive association with error under our within-arm directional tests. Appendix~\ref{sec:invariants} separately reports a small invariant-guidance pilot; this pilot does not establish either an accuracy benefit or an efficiency benefit.

This diagnostic is relevant to verifier-facing work because counterexample replay, symbolic-execution guidance, invariant assessment, and diagnosis of failed proof obligations involve values at specific program points. The benchmark evaluates a model's ability to recover those values from code. It measures one component of these workflows. Software correctness still requires a specification and independent verification.

\section{Related Work}

\paragraph{Execution prediction.}
CRUXEval is the main precedent for this work~\citep{cruxeval}. It tests final input and output behavior on short generated functions. CRUXEval-X extends those tasks to 19 programming languages~\citep{cruxevalx}. LiveCodeBench includes an execution subtask and supplies the contest programs behind our two largest studies~\citep{livecodebench}. REval adds runtime probes over short functions, including coverage, one variable after a selected line, the next statement, and final output~\citep{reval}. Our benchmark contributes paired execution scales and joint projections of core state at two deterministic checkpoints over substantially larger programs. Its sources include contest programs, authored algorithms, and one library implementation.

\paragraph{Learning from execution traces.}
Learning to Execute studies neural interpretation of synthetic programs~\citep{learningtoexecute}. CodeMind and other code-simulation evaluations document fragility on iterative execution tasks~\citep{codemind,lamalfa2024}. CodeExecutor trains a model to emit complete execution traces~\citep{codeexecutor}, while NExT uses execution-grounded rationales to improve code reasoning and repair~\citep{next}. Scratchpads and chain-of-thought prompting show that allocating tokens to intermediate computation can improve program evaluation~\citep{scratchpad,cot}. CWM adds interpreter and environment trajectories during training and reports strong explicit trace prediction and CRUXEval performance~\citep{cwm}. Our study takes the complementary evaluation view. It removes runtime feedback, varies the demanded execution, and grades only the declared endpoint or checkpoint.

\paragraph{Coding harnesses.}
Repository-level coding systems combine a language model with context management, editing operations, shell access, tests, and recovery logic. SWE-agent makes interface design an experimental variable~\citep{sweagent}. OpenHands provides a model-agnostic runtime and tool platform~\citep{openhands}. Agentless shows that a fixed three-stage workflow can compete with more autonomous designs~\citep{agentless}. Agent evaluations measure whether this combined system completes a software task. Model-only execution prediction measures whether a served model can determine fixed program behavior without delegating computation to those components.

\paragraph{Invariants.}
Program invariants summarize relations preserved across repeated execution and support inductive proofs. Prior work trains language models to infer invariants for program analysis and verification~\citep{pei2023}. Our pilot asks the complementary question of whether supplying a source-checked, answer-free invariant helps exact output prediction. This connects the execution diagnostic to a possible use of semantic summaries during reasoning.

\section{Benchmark Construction}

The benchmark turns these execution questions into four related tasks per case. Its construction first fixes the program, inputs, and state projection, then checks the resulting tasks by execution before collecting any model responses.

\subsection{Task unit and four arms}

A case begins with a deterministic program, two legal inputs, one selected loop, and an ordered projection of live variables. A task presents exact source and input bytes and requests either the program's final standard output or the projected state at a declared checkpoint. Ground-truth files are generated by execution and never enter the model context. Table~\ref{tab:arms} defines the four arms.

\begin{table}[h]
\centering
\small
\begin{tabular}{@{}lll@{}}
\toprule
\textbf{Arm} & \textbf{Program and input} & \textbf{Target} \\
\midrule
\texttt{short-trace-final} & clean source, shorter-trace input & final output \\
\texttt{long-trace-final}  & same source, longer-trace input   & final output \\
\texttt{inside-loop-state} & instrumented source, long input & state inside loop \\
\texttt{post-loop-state}   & instrumented source, long input & state after loop \\
\bottomrule
\end{tabular}
\caption{The canonical four-arm design. The two state arms emit compact JSON with the same ordered fields and top-level types. At least one core field must change between checkpoints.}
\label{tab:arms}
\end{table}

The two final-output arms share identical source bytes and differ in input. Their labels record the measured ordering of native trace lengths. This pairing controls source identity, while branch paths, arithmetic, intermediate values, and answer length may also change. The state arms expose the same projection at two points in the long-input execution. The inside arm emits state at a deterministic iteration before the selected loop body continues, then terminates. The post arm lets that loop finish normally, emits the projection immediately afterward, and terminates before later code can alter it. Comparing these arms with final output bundles instrumented source, checkpoint semantics, target type, and JSON formatting. Appendix~\ref{app:construction-example} shows the transformations for Figure~\ref{fig:running-example}.

\subsection{Source pools and selection}

The benchmark draws from four source pools. The two LiveCodeBench studies inspect all 350 Hard programs at a pinned revision and initially retain 320. A final construction audit excludes two Python programs with nondeterministic outputs, leaving 283 Python and 35 C++ cases. The CodeContests study~\citep{alphacode} begins with 282 programs rated between 2400 and 3500. Seventy-nine meet the language, input, and cohort requirements, and repeated execution plus checkpoint checks retain 22. The authored study covers 30 textbook algorithms across dynamic programming, graph algorithms, strings, sorting, numerical methods, geometry, and simulation. The NetworkX study~\citep{networkx} holds one network-simplex implementation fixed and selects 30 graph inputs from 90,000 generated candidates in three predeclared runtime bands. We fix selection and exclusion without using any of the models' accuracy results.

For each candidate, a deterministic builder prepares legal short and long inputs, identifies a stable source anchor for the selected loop, and constructs both state projections. Admission requires a strict short-trace $<$ long-trace ordering in three repeated measurements. The builder also checks checkpoint reachability, type and field agreement, a changed core field between checkpoints, deterministic oracles, and preservation of the clean program's output under instrumentation. A later audit corrected checkpoint identities for seven CodeContests cases. The 69 affected state predictions were invalidated and recollected under the corrected task identities.

The admitted cases therefore have reproducible execution targets under custom inputs and checkpoints. Figure~\ref{fig:running-example} and Appendix~\ref{app:construction-example} illustrate the inputs, checkpoint transformations, and oracle values for an admitted case.

\subsection{Corpus statistics}

Table~\ref{tab:corpus} reports the resulting corpus. The cyclomatic-complexity summary uses the clean source shown in the short arm and language-specific control-flow conventions. The values describe variation within each cohort, with the largest cohort spanning 3 to 212. NetworkX adds execution diversity through 30 inputs for one fixed routine. This combination lets us examine both variation across programs and variation in the execution of a shared implementation.

\begin{table}[h]
\centering
\small
\setlength{\tabcolsep}{4.5pt}
\begin{tabular}{@{}llrrrl@{}}
\toprule
\textbf{Study} & \textbf{Lang.} & \textbf{Cases} & \textbf{Sources} & \textbf{Graded} & \textbf{CC median [range]} \\
\midrule
\lcbpy{}   & Python & 283 & 283 & 7{,}881 & 16 [3, 212] \\
\lcbcpp{}  & C++20  & 35  & 35  & 978       & 20 [6, 95] \\
\classic{} & Python & 30  & 30  & 840       & 7.5 [5, 13] \\
\cc{}      & C++    & 22  & 22  & 612       & 39.5 [13, 128] \\
\nx{}      & Python & 30  & 1   & 840       & 88 [88, 88] \\
\midrule
\textbf{Total} & & \textbf{400} & \textbf{371} & \textbf{11{,}151} & \\
\bottomrule
\end{tabular}
\caption{Benchmark composition. Graded counts completed predictions included in accuracy, out of 11,200 planned. CC is \Occ{} on the clean short-arm source. The NetworkX value repeats because its cases share one source.}
\label{tab:corpus}
\end{table}

\section{Evaluation}

With the tasks and oracles fixed, we measure how accuracy changes with the requested reasoning mode and across the four task conditions. Each study uses a fixed collection protocol and grading rule throughout its main grid.

\subsection{Models, sessions, and grading}

The model matrix has seven settings from four model families. We evaluate \texttt{gpt-5.6-sol} at \texttt{high} and \texttt{off}, \texttt{glm-5.3} at \texttt{high}, \texttt{deepseek-v4-pro-0813} at \texttt{high} and \texttt{off}, and \texttt{qwen3.8-27b} at \texttt{high} and \texttt{off}. These are provider-specific requests. The \texttt{gpt-5.6-sol-off} configuration maps our off label to the provider's \texttt{minimal} reasoning value. The other off settings use the corresponding provider's no-reasoning value. High-reasoning requests have a 3,600-second timeout and off requests have a 1,800-second timeout. The experiment files record no temperature or top-$p$ override. Each task runs in a fresh session with tools, retrieval, and code execution disabled. We retain one selected response per model--case--arm cell and do not retry completed wrong answers. This protocol reports selected-answer accuracy; no repeated-query trials estimate its stability. Appendix~\ref{app:grading} gives the prompt template and response contract.

Correctness follows committed study-specific oracle-equivalence rules. Parsed JSON values are compared structurally where applicable, most text outputs are compared as token sequences, and NetworkX uses byte equality. We report both completed-response accuracy, $C/R$, and completion-aware accuracy, $C/N$, where $C$ is correct, $R$ completed, and $N$ planned. Missing responses count as wrong in $C/N$; completed wrong answers count against both scores. Correct answers count regardless of response-envelope validity. All settings plan 1,600 predictions, or 400 per arm. Forty-nine requests return no response: 47 for \texttt{gpt-5.6-sol-off} and two for \texttt{qwen3.8-27b-high}. Their overall $C/R$ and $C/N$ scores are 25.2\%/24.5\% and 39.4\%/39.3\%; the setting ordering is unchanged. Appendix~\ref{app:grading} gives grading and envelope details.

\subsection{Overall results}

The largest accuracy differences in the model matrix are between reasoning modes. Across all studies, \texttt{gpt-5.6-sol} scores 74.8\% with high reasoning and 25.2\% with reasoning off on completed responses, or 74.8\% and 24.5\% on planned responses. The high/off scores are 63.8\%/8.6\% under both denominators for \texttt{deepseek-v4-pro-0813}, and 39.4\%/6.3\% completed or 39.3\%/6.3\% planned for \texttt{qwen3.8-27b}. On the fully gradable classic grid, with 120 predictions per setting, the three pairs are 82.5\% versus 20.0\%, 83.3\% versus 8.3\%, and 30.0\% versus 3.3\%. These compare each provider's reasoning settings, whose meanings and computational budgets vary by provider.

\begin{table}[h]
\centering\small
\setlength{\tabcolsep}{3.5pt}
\begin{tabular}{@{}lrrrr@{}}
\toprule
Setting & Short final & Long final & Inside state & Post state \\
\midrule
\texttt{gpt-5.6-sol-high} & 93.0/93.0 & 77.0/77.0 & 65.5/65.5 & 63.5/63.5 \\
\texttt{gpt-5.6-sol-off} & 46.2/46.2 & 32.9/32.5 & 7.8/7.5 & 12.6/11.8 \\
\texttt{glm-5.3-high} & 87.2/87.2 & 64.5/64.5 & 36.2/36.2 & 33.2/33.2 \\
\texttt{deepseek-v4-high} & 91.5/91.5 & 71.0/71.0 & 52.8/52.8 & 39.8/39.8 \\
\texttt{deepseek-v4-off} & 19.2/19.2 & 14.5/14.5 & 0.2/0.2 & 0.5/0.5 \\
\texttt{qwen3.8-27b-high} & 82.0/82.0 & 44.2/44.2 & 16.5/16.5 & 14.5/14.5 \\
\texttt{qwen3.8-27b-off} & 12.2/12.2 & 13.0/13.0 & 0.0/0.0 & 0.0/0.0 \\
\bottomrule
\end{tabular}
\caption{Accuracy by arm, in percent. Each cell gives completed-response accuracy $C/R$ followed by all-planned accuracy $C/N$, which counts nonresponses as wrong; $N=400$ per cell. Both valid- and invalid-envelope correct answers count. The model label \texttt{deepseek-v4-pro-0813} is shortened to \texttt{deepseek-v4} in this table.}
\label{tab:arm-accuracy}
\end{table}

The reasoning advantage persists across task conditions, but accuracy also drops substantially on the longer-input and state tasks. For the strongest setting, Table~\ref{tab:arm-accuracy} shows a decline from 93.0\% on short final output to 77.0\% on long final output and 65.5\% and 63.5\% on checkpoint state. All four high-reasoning settings follow this ordering from short output to long output to state prediction. The benchmark thus exposes errors that short-output scores alone would leave unseen. These are task-condition comparisons. Changing the input can alter branches and answer properties, while state tasks add instrumentation, a checkpoint, and a structured target.

The fixed-implementation NetworkX study exposes a particularly difficult set of executions. The best setting answers 6 of 120 tasks and none of its 60 state tasks. The other six settings answer none of 120. This floor leaves little resolution for ranking settings or explaining errors, and one routine does not represent library code generally. More library routines at intermediate difficulty are needed. Appendix~\ref{app:accuracy-detail} gives the study-by-setting matrix with both denominators.

\section{Quantitative Analysis}
\label{sec:quantitative}

The accuracy results locate the conditions under which predictions fail. We next examine how those failures relate to properties of the source and its execution, using one static and three dynamic measurements. Cyclomatic complexity is \Occ{}$=E-N+2$ on exact source bytes. Native trace length \Otrace{} counts adapter-defined execution events. Peak state size is \Osize{}$=\max_t |S_t|$. Cumulative state load is
\begin{equation}
\Oload{}=L=\sum_{t=1}^{T}|S_t|,
\label{eq:stateload}
\end{equation}
where $S_t$ is the reachable state owned by the target program at observation $t$. Summing state size over the execution makes $L$ capture both the number of observed steps and how much state is present at each of those steps.

Python measurements use CPython opcode events and reachable-state walks at declared mutation boundaries. C++ measurements use Clang coverage-region counts and a separate state convention. We analyze their raw values separately. The original analysis uses a normalized Mann--Whitney statistic $\theta>0.5$ to indicate larger metric values among failures, with a one-sided permutation test from 10,000 fixed-seed shuffles. Its univariate logistic slopes describe error odds per metric doubling. Appendix~\ref{app:analysis-detail} preserves these statistics and their caveats about dependence; the adjusted analysis below uses uncertainty estimates clustered by source program.

\subsection{Paired shorter- and longer-input results}

We first compare the paired final-output tasks, which hold source byte-identical. Across 2,397 matched Python case--setting pairs, moving from the selected shorter-trace input to the longer-trace input yields 528 correct-to-wrong changes and 147 wrong-to-correct changes. Among the four high-reasoning settings, the counts are 339 and 26. Correct-to-wrong changes are therefore much more frequent on the selected longer inputs. As in the arm-level comparison, this asymmetry includes changes in branch paths, values, and output properties as well as trace length.

A post-review sensitivity resamples whole clean-source clusters, preserving all matched settings and arms for each selected source. The paired short-minus-long accuracy difference is 15.89 percentage points (95\% percentile interval [12.12, 20.10]; 10,000 replicates, 314 source clusters). The 30 NetworkX cases share one source and therefore form one cluster. Excluding them gives 17.33 points ([13.95, 20.81]; 2,187 pairs from 313 sources). These intervals describe variation among admitted sources under the selected inputs and settings; they do not isolate a causal effect of trace length.

\subsection{Metric associations after adjustment}

The original pooled analyses combine four task conditions and treat their rows as independent. They are retained as descriptive comparisons in Appendix~\ref{app:original-pooled}; they do not support inferential claims about a failure mechanism. We therefore add a post-review analysis of the Python LiveCodeBench and classic cohorts, adjusting for study, arm, and model setting and clustering uncertainty by the clean source program~\citep{cameronmiller2015}. NetworkX is excluded from these adjusted fits because it has one source and near-floor outcomes. The C++ analysis remains separate because its dynamic measurements use an adapter that differs from the one used for Python.

A post-review exploratory logistic analysis conditions on study, task arm, and model-setting intercepts and clusters uncertainty by clean source. It uses the 8,721 completed Python LiveCodeBench and classic predictions from 313 distinct sources; NetworkX is excluded because its 30 inputs reuse one implementation. In a joint model of log-base-two trace length and peak state size, neither coefficient has a source-cluster 95\% interval wholly above zero. Adding cumulative state load does not reveal a positive incremental association: its error odds ratio per doubling is 0.859 (95\% interval [0.733, 1.006]). The three dynamic predictors are strongly correlated, so conditional coefficients require care. The pooled plots remain descriptive; they do not show that cumulative state load independently predicts error beyond its components. The full results, together with the numerical checks, are reported in Appendix~\ref{app:cluster-analysis}.

Cumulative state load can describe observed executions without being a unique explanation for error. Models may simulate updates, derive a closed form, or recognize an algorithm. Accuracy and metric associations cannot distinguish these strategies or identify the model's internal computation.

\subsection{Cyclomatic complexity within task conditions}

Dynamic measurements describe a particular execution. We now examine cyclomatic complexity, a source-level measure of control-flow structure. We test \Occ{} separately for every model, study, and arm because instrumentation changes the state-arm source. Ninety-four of 140 series contain both successes and failures. Twelve meet the unadjusted nominal rule, including seven on \texttt{post-loop-state}, three on \texttt{inside-loop-state}, and two on \texttt{short-trace-final}. None of 29 estimable \texttt{long-trace-final} series meets the rule. This clean-source arm avoids instrumentation differences and has lower ceiling effects than the short arm.

Higher cyclomatic complexity therefore shows no consistent positive association with error within the tested conditions. This directional result concerns CC alone. It leaves open negative or nonmonotonic associations and the predictive value of other static properties, such as source length, nesting, and data-flow structure.

\section{Limitations and Future Work}

The results establish accuracy differences across selected task conditions. We have not measured the fraction of short/long pairs with preserved branch behavior or run controls that vary iteration count alone. Even when branch structure, output type, and format agree, values, arithmetic burden, and output length may differ. Future construction should preserve branch decisions while scaling repeated work and independently vary state width, checkpoint position, and answer format.

The source-clustered reanalysis accounts for related predictions and measured task covariates; it cannot remove unmeasured input or semantic confounding. It is exploratory and was added after review. Dynamic measurements depend on adapters and instrumentation, so raw Python and C++ values are not interchangeable. Multiplicity and predictor correlation limit conclusions from the original metric tests. Broader static analysis should include source length, nesting, and data flow alongside cyclomatic complexity.

Coverage is concentrated in contest programs and selected algorithms. Admission favors explicit loops with deterministic, reachable checkpoints and changing state projections. It excludes some recursive, comprehension-based, short-loop, and large-state programs, so the corpus does not cover all forms of execution reasoning. Custom inputs do not rule out familiarity with public source or algorithms~\citep{riddell2024}; contamination is unquantified. The single NetworkX routine produces floor effects and needs intermediate-difficulty library counterparts. Public release also permits benchmark-specific tuning; future evaluation should use held-out sources and inputs.

The main grid retains one selected response per cell, without a repeated-query stability analysis. Resampling source programs estimates uncertainty across this corpus, not variation from repeated API calls. Provider-specific reasoning modes have unequal and incompletely observed compute budgets, and endpoints may drift. Provider finish reasons, token-level accounting, and native sessions are unavailable for this analysis, so it cannot identify truncation or recover sampling variance. Grading statuses distinguish envelope compliance and correctness but do not classify a wrong answer as an execution, arithmetic, copying, or formatting mistake. That taxonomy needs separately audited annotations.

The invariant pilot is separate from the main grid and establishes no reliable accuracy or efficiency benefit. Its exact prompts and native sessions are not reproduced here. Appendix~\ref{sec:invariants} records the available aggregate evidence and the controls needed for a confirmatory study.

\section{Conclusion}

Exact output and checkpoint-state prediction expose gaps in model-only code reasoning across 400 cases and four related task conditions. Reasoning-enabled settings perform much better than their off counterparts, yet the strongest setting falls from 93.0\% on shorter-input final output to 77.0\% on longer-input final output and below 66\% on both state tasks. Completion-aware scoring preserves this pattern. Paired comparisons hold source fixed, and source-clustered analyses account for related predictions, while changed inputs and checkpoint targets remain bundled interventions. The benchmark provides concrete tasks and execution-generated oracles for studying these differences without treating them as proof of a particular internal failure mechanism.

\begin{ack}
\FundingDisclosure
\end{ack}

\appendix

\section{Protocol and Analysis Scope}
\label{app:reproducibility}

The main text specifies the four task arms, source-pool selection, execution checks, model settings, and scoring denominators. Appendix~\ref{app:grading} gives the prompt and grading rules; Appendix~\ref{app:accuracy-detail} gives the complete accuracy matrix; Appendix~\ref{app:cluster-analysis} specifies the source-clustered regression and bootstrap procedures. Appendix~\ref{sec:invariants} reports the separate invariant pilot, and Appendix~\ref{app:construction-example} gives the checkpoint transformations and oracle values for the running example.

The reported evidence includes 11,151 completed predictions and 49 no-response records across 11,200 planned cells. Case-level tasks and responses and raw execution traces are not reproduced in this paper, so the aggregate tables alone do not support independent regrading or recomputation of the execution measurements.

\section{Grading Detail}
\label{app:grading}

Correctness and response-envelope validity produce four statuses. Both correct statuses count toward accuracy. The \classic{} and LiveCodeBench studies compare parsed JSON values when the oracle is JSON and otherwise compare whitespace-separated tokens. \cc{} compares ordered signed-integer and alphabetic-token sequences while ignoring separator punctuation. \nx{} compares the answer and oracle byte-for-byte. These rules produce 4,361 correct predictions. Outer-whitespace normalization alone would reject 12 LiveCodeBench Python answers accepted by structural JSON equality and accept five NetworkX answers rejected by byte equality. The reported results retain the committed within-study policies.

The rendered prompt supplies literal source and input in \texttt{<program>} and \texttt{<input>} blocks. The requested outer envelope is a JSON object with one string field, \texttt{output}, whose contents are the exact predicted text. A state answer is therefore serialized JSON inside that string. The system prompt uses the template below, substituting the literal task source and input for the two placeholders. The user message is ``Return the requested prediction.'' There are no per-task prompt overrides.

\begin{quote}
\begin{minipage}{\linewidth}
\small
\begin{verbatim}
Given the program and input below, determine the exact output.

<program>
[program source]
</program>

<input>
[input text]
</input>

Respond in this format:

{"output":"<exact text>"}
\end{verbatim}
\end{minipage}
\end{quote}

A completed wrong answer counts against the model regardless of envelope validity. A completed response without a final answer is wrong rather than excluded, including five \nx{} responses that contain reasoning without an answer. Only transport failures and requests returning no response are excluded. Of 49 excluded requests, 47 are \texttt{gpt-5.6-sol-off} and two are \texttt{qwen3.8-27b-high}.

Of 11,151 completed responses, 1,231 (11.0\%) violate the required outer response envelope: 276 remain correct under conservative answer extraction and 955 do not match the oracle. The 5,835 wrong-valid responses follow the envelope but contain a wrong answer; the remaining 4,085 are correct-valid. Table~\ref{tab:completion-envelope} shows this breakdown by setting alongside the all-planned sensitivity that counts nonresponses as wrong. Envelope validity concerns only the outer single-string \texttt{output} object, not the validity or semantic accuracy of a JSON state inside that string. These observable outcomes cannot distinguish arithmetic, state-transition, input-reading, copying, or provider-budget failures. Native sessions, stop reasons, token usage, and effective provider output limits are unavailable for this analysis, so it cannot recover a truncation rate or an equal-budget comparison.

\begin{table}[h]
\centering\scriptsize
\begin{tabular}{@{}lrrrr@{}}
\toprule
Setting & Completed & Invalid envelope & Accuracy & All planned \\
\midrule
\texttt{deepseek-v4-pro-0813-high} & 1600 & 133 & 63.8\% & 63.8\% \\
\texttt{deepseek-v4-pro-0813-off} & 1600 & 154 & 8.6\% & 8.6\% \\
\texttt{glm-5.3-high} & 1600 & 396 & 55.3\% & 55.3\% \\
\texttt{gpt-5.6-sol-high} & 1600 & 18 & 74.8\% & 74.8\% \\
\texttt{gpt-5.6-sol-off} & 1553 & 55 & 25.2\% & 24.5\% \\
\texttt{qwen3.8-27b-high} & 1598 & 429 & 39.4\% & 39.3\% \\
\texttt{qwen3.8-27b-off} & 1600 & 46 & 6.3\% & 6.3\% \\
\bottomrule
\end{tabular}
\caption{Completed-response accuracy, invalid outer envelopes, and the all-planned sensitivity. Each setting plans 1,600 cells. All-planned accuracy counts no responses as wrong.}
\label{tab:completion-envelope}
\end{table}

\FloatBarrier
\Needspace{28\baselineskip}
\section{Complete Accuracy Matrix}
\label{app:accuracy-detail}

\begin{table}[h]
\centering\small
\setlength{\tabcolsep}{4pt}
\textbf{Completed-response accuracy ($C/R$)}\\[3pt]
\begin{tabular}{@{}lrrrrr@{}}
\toprule
Setting & \texttt{lcb\_python} & \texttt{lcb\_cpp} & \texttt{classic} & \texttt{codecontests} & \texttt{networkx} \\
\midrule
\texttt{gpt-5.6-sol-high} & 80.7 & 75.7 & 82.5 & 81.8 & 5.0 \\
\texttt{gpt-5.6-sol-off} & 29.8 & 22.5 & 20.0 & 15.1 & 0.0 \\
\texttt{glm-5.3-high} & 60.0 & 60.0 & 56.7 & 61.4 & 0.0 \\
\texttt{deepseek-v4-high} & 65.9 & 72.9 & 83.3 & 81.8 & 0.0 \\
\texttt{deepseek-v4-off} & 10.2 & 5.0 & 8.3 & 6.8 & 0.0 \\
\texttt{qwen3.8-27b-high} & 43.4 & 44.3 & 30.0 & 46.5 & 0.0 \\
\texttt{qwen3.8-27b-off} & 7.4 & 4.3 & 3.3 & 8.0 & 0.0 \\
\bottomrule
\end{tabular}
\\[9pt]
\textbf{All-planned accuracy ($C/N$)}\\[3pt]
\begin{tabular}{@{}lrrrrr@{}}
\toprule
Setting & \texttt{lcb\_python} & \texttt{lcb\_cpp} & \texttt{classic} & \texttt{codecontests} & \texttt{networkx} \\
\midrule
\texttt{gpt-5.6-sol-high} & 80.7 & 75.7 & 82.5 & 81.8 & 5.0 \\
\texttt{gpt-5.6-sol-off} & 28.6 & 22.1 & 20.0 & 14.8 & 0.0 \\
\texttt{glm-5.3-high} & 60.0 & 60.0 & 56.7 & 61.4 & 0.0 \\
\texttt{deepseek-v4-high} & 65.9 & 72.9 & 83.3 & 81.8 & 0.0 \\
\texttt{deepseek-v4-off} & 10.2 & 5.0 & 8.3 & 6.8 & 0.0 \\
\texttt{qwen3.8-27b-high} & 43.4 & 44.3 & 30.0 & 45.5 & 0.0 \\
\texttt{qwen3.8-27b-off} & 7.4 & 4.3 & 3.3 & 8.0 & 0.0 \\
\bottomrule
\end{tabular}
\caption{Overall accuracy by study and setting, in percent. The upper matrix excludes nonresponses from the denominator; the lower counts them as wrong. Planned denominators per setting are 1,132, 140, 120, 88, and 120, respectively. DeepSeek labels abbreviate \texttt{deepseek-v4-pro-0813}.}
\label{tab:accuracy}
\end{table}

\FloatBarrier
\section{Source-Clustered Reanalysis}
\label{app:cluster-analysis}

The post-review regression treats a completed response as one observation and its incorrectness as the binary outcome. The baseline has a study intercept, three arm indicators, and six setting indicators; reference levels are \classic{}, \texttt{short-trace-final}, and \texttt{deepseek-v4-pro-0813-high}. The second model adds mean-centered $\log_2$ native trace length and peak state size; the third adds mean-centered $\log_2$ cumulative state load. Slopes therefore measure change in error log odds for a doubling of one predictor conditional on the others. All observations have complete normalized measurements. The 30 NetworkX inputs reuse one implementation and display a strong accuracy floor; they are excluded, as are the C++ cohorts whose adapters measure different quantities.

Maximum-likelihood estimates use damped Newton steps. The sandwich covariance sums score vectors across every arm and setting of each clean-source SHA-256 cluster, with the finite-sample factor $G/(G-1)\,(N-1)/(N-p)$. Intervals and two-sided Wald tests use a $t_{G-1}$ reference, where $G=313$ and $N=8{,}721$. All three fits converge in six iterations; their design matrices have full column rank, information-matrix condition numbers range from 139 to 4,337, maximum absolute score residuals are below $1.3\times10^{-10}$, and covariance matrices are positive definite.

Table~\ref{tab:cluster-logistic} reports the dynamic coefficients. Correlations between the log predictors are 0.754 (trace--peak), 0.932 (trace--load), and 0.923 (peak--load). Conditional trace and peak coefficients increase when load is added, while the load estimate is negative and its interval includes one on the odds-ratio scale. This sign pattern does not support a positive incremental role for the composite. Model misspecification, collinearity, selected inputs, and omitted answer properties remain possible explanations. No coefficient establishes causal effects or out-of-source predictive improvement. These analyses were added after review, are exploratory, and use no multiplicity adjustment.

\begin{table}[h]
\centering\small
\begin{tabular}{@{}llrrr@{}}
\toprule
Model & Predictor (per doubling) & Odds ratio & 95\% interval & $p$ \\
\midrule
Trace + peak & Trace & 1.049 & [0.988, 1.113] & 0.118 \\
Trace + peak & Peak state & 1.024 & [0.964, 1.087] & 0.447 \\
Trace + peak + load & Trace & 1.205 & [1.025, 1.418] & 0.024 \\
Trace + peak + load & Peak state & 1.183 & [1.009, 1.387] & 0.039 \\
Trace + peak + load & State load & 0.859 & [0.733, 1.006] & 0.059 \\
\bottomrule
\end{tabular}
\caption{Exploratory logistic models for prediction error. Both models include study, arm, and setting intercepts; 95\% intervals and $p$-values use source-cluster uncertainty. Odds ratios are per doubling of the named metric.}
\label{tab:cluster-logistic}
\end{table}

For the paired comparison, 10,000 bootstrap replicates sample whole clean-source clusters with replacement, keeping all retained case--setting pairs together. Each replicate divides the summed short-minus-long correctness by its summed pair count. Resampling is pooled across studies; the NetworkX source, when sampled, carries all 30 inputs. The reported percentile intervals condition on the admitted source set and selected model settings. They quantify neither training-data contamination nor generalization to a representative program population.

\begin{table}[h]
\centering\small
\begin{tabular}{@{}llrrrr@{}}
\toprule
Cohort & Settings & Pairs & Sources & Decrement (pp) & 95\% interval \\
\midrule
All Python & All seven & 2397 & 314 & 15.89 & [12.12, 20.10] \\
All Python & Four high & 1372 & 314 & 22.81 & [18.02, 27.92] \\
Exclude NetworkX & All seven & 2187 & 313 & 17.33 & [13.95, 20.81] \\
Exclude NetworkX & Four high & 1252 & 313 & 24.84 & [20.93, 28.83] \\
\bottomrule
\end{tabular}
\caption{Paired short-minus-long decrement with source-cluster percentile intervals. The NetworkX exclusion sensitivity removes the single source reused for its 30 cases.}
\label{tab:paired-bootstrap}
\end{table}

\section{Original Exploratory Metric Analyses}
\label{app:analysis-detail}
\label{app:original-pooled}

These analyses reproduce the submitted results. Their row-level permutation tests and Wald intervals do not account for repeated arms from the same case. We retain their statistics for comparison, without treating nominal significance as confirmatory evidence.

To examine the broader relation with execution scale, we pool the four arms within each setting and study. Peak state size and cumulative state load meet the nominal association rule in 14 of 15 estimable Python series, and trace length does so in 12. Under the C++ adapter the corresponding counts are 0 of 14, 4 of 14, and 9 of 14. The association with cumulative state load is therefore more consistent across the Python series, where execution length and reachable state are measured under a shared convention.

\begin{figure}[t]
\centering
\includegraphics[width=\textwidth]{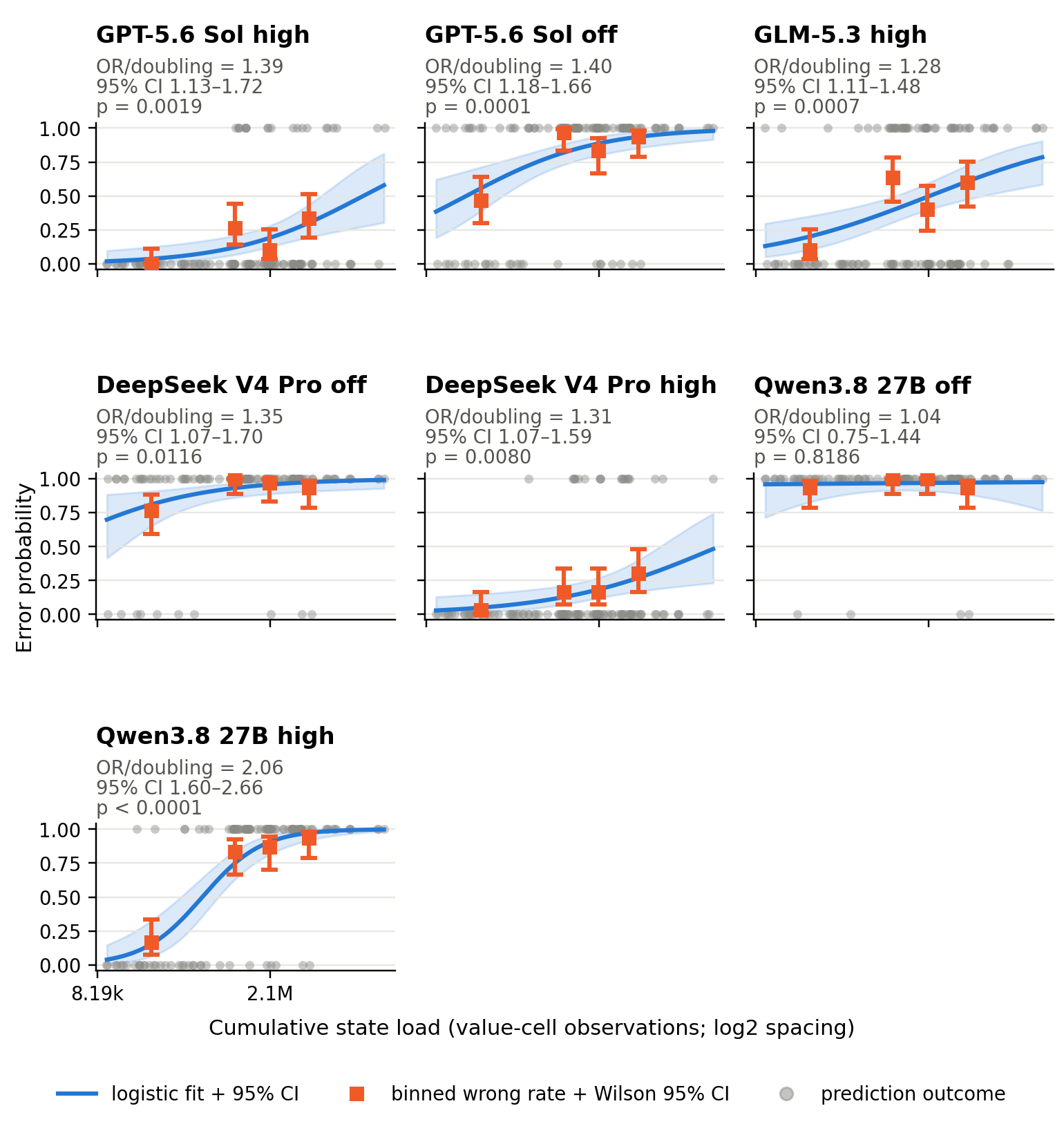}
\caption{Error probability against cumulative state load on the \classic{} study, one panel per setting. Squares are binned error rates. Curves and bands show descriptive logistic fits and 95\% Wald intervals. These intervals and the panel-header $p$-values are not cluster robust.}
\label{fig:stateload}
\end{figure}

\begin{figure}[t]
\centering
\includegraphics[width=\textwidth]{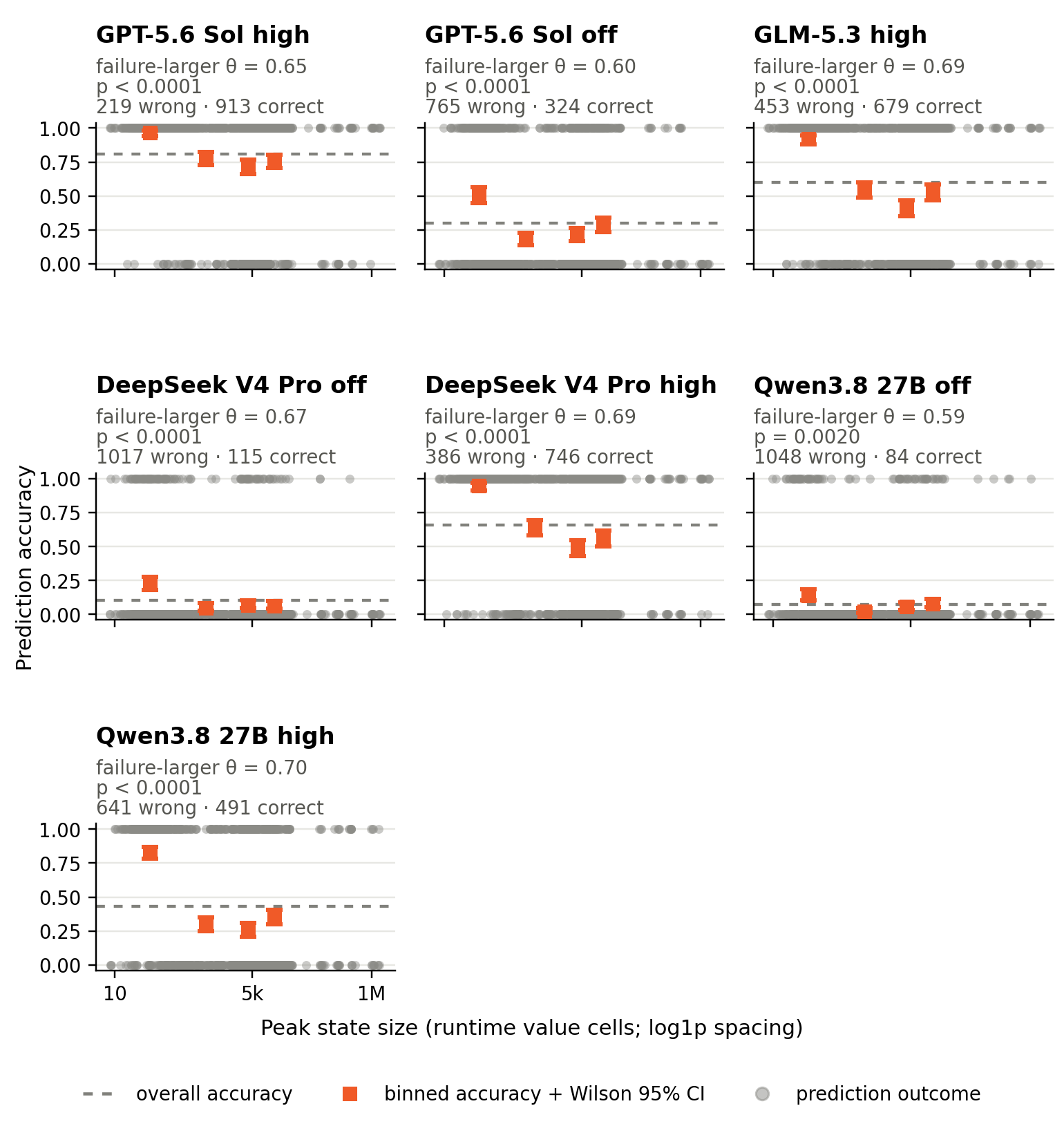}
\caption{Accuracy against peak state size on \lcbpy{}, with all four arms pooled. Points are predictions, squares are binned accuracy with Wilson intervals, and dashed lines mark overall accuracy. Panel headers report the predeclared rank statistic and nominal permutation $p$-value.}
\label{fig:statesize}
\end{figure}

On \lcbpy{}, pooled odds ratios per doubling of cumulative state load range from 1.05 to 1.14, and all seven slopes are nominally significant. On \classic{}, six of seven finite fits are nominally significant, with supported odds ratios from 1.28 to 2.06, as shown in Figure~\ref{fig:stateload}. The only finite NetworkX fit is 3.28 with a 95\% interval from 1.28 to 8.40, based on six correct predictions. The fourteen C++ fits range from 1.00 to 1.09, with three nominally significant. These slopes are specific to each cohort and measurement adapter.

These pooled associations are exploratory because each case contributes four related rows and arm identity is unmodeled. The conventional Wald intervals and row-level permutation $p$-values do not account for that dependence. Within-arm evidence is much weaker. Of 420 candidate dynamic series, 282 are estimable and 41 meet the raw nominal rule. The remaining 138 have the same outcome for every response. None of the 45 estimable Python long-final tests in the declared multiplicity family survives Benjamini--Hochberg correction at 0.05. The supported conclusion is a pooled descriptive association whose strength depends on the task grouping and measurement convention.

One possible explanation is that models maintain or reconstruct concrete state through a sequence of updates. Longer executions offer more opportunities for transition errors, while wider states require more values to remain precise. Equation~\ref{eq:stateload} represents both demands. This is a hypothesis about internal computation, which output accuracy alone cannot resolve. Models may also use closed forms, familiar algorithms, or other shortcuts. Trace length, peak state size, and cumulative state load are correlated, and the present analysis does not establish whether the composite adds predictive value beyond its components.

For a series of $(x,y)$ points with $y=0$ failure and $y=1$ success, the normalized rank statistic is
\[
\theta = \frac{1}{n_f n_s}\sum_{f,s}\left[\mathbf{1}(x_f>x_s)+\tfrac12\mathbf{1}(x_f=x_s)\right].
\]
The one-sided permutation test shuffles outcome labels 10,000 times under a fixed seed. All-correct and all-wrong series are inestimable. Static metrics are tested per arm because state instrumentation changes source bytes. The dynamic headline rows pool all four arms for each model--study setting, while the corrected family is restricted to the 45 estimable Python long-final tests.

\begin{table}[h]
\centering
\small
\begin{tabular}{@{}lrrrr@{}}
\toprule
& \multicolumn{2}{c}{\textbf{Python cohorts}} & \multicolumn{2}{c}{\textbf{C++ cohorts}} \\
\cmidrule(lr){2-3}\cmidrule(lr){4-5}
\textbf{Metric} & \textbf{Supported} & \textbf{$\theta$ range} & \textbf{Supported} & \textbf{$\theta$ range} \\
\midrule
\Osize{}  & 14 / 15 & 0.51--0.87 & 0 / 14 & 0.32--0.64 \\
\Oload{}  & 14 / 15 & 0.48--0.88 & 4 / 14 & 0.52--0.70 \\
\Otrace{} & 12 / 15 & 0.51--0.89 & 9 / 14 & 0.44--0.80 \\
\bottomrule
\end{tabular}
\caption{Pooled support for dynamic metrics by adapter. A supported row has $\theta>0.5$ and unadjusted $p<0.05$. The six inestimable Python settings are all-wrong NetworkX settings. Ranges span all estimable series and are not comparable across language groups.}
\label{tab:dynamic}
\end{table}

\FloatBarrier
\section{Exploratory Invariant-Guidance Pilot}
\label{sec:invariants}

The benchmark motivates a separate practical question. A relation between initial and current state can sometimes replace repeated computation when deriving an endpoint. We use \emph{context compression} for this proposed role of an invariant and explore it in a separate pilot.

The pilot covers 12 difficult long-trace programs, six in Python and six in C++. GPT-5.6 Sol at high reasoning answered each case in three fresh sessions under a baseline prompt and a prompt containing one audited invariant. The invariant stated a recurrence, run-length property, or decomposition rule. Each was checked against the source using small examples, an independent calculation, or direct execution. None was formally proved. The guidance included neither the final output nor an evaluated endpoint. We repeated the comparison under two response contracts, giving 144 predictions in total.

\begin{table}[h]
\centering
\small
\begin{tabular}{@{}lrrr@{}}
\toprule
\textbf{Response contract} & \textbf{Baseline} & \textbf{Invariant} & \textbf{Difference} \\
\midrule
Exact output only & 24 / 36 & 26 / 36 & +2 \\
Visible rationale, then output & 24 / 36 & 24 / 36 & 0 \\
\bottomrule
\end{tabular}
\caption{Exploratory invariant pilot, outside the 400-case main grid. Each cell contains 12 programs times three fresh sessions. Only the exact output is graded.}
\label{tab:invariant}
\end{table}

Table~\ref{tab:invariant} shows a two-answer gain under the exact-output contract and no aggregate gain under the visible-rationale contract. One case moves from 0/3 to 1/3 in both comparisons, while most remain unchanged. Rationale inspection finds answers that state the supplied recurrence correctly and then lose precision in arithmetic, state tracking, input reading, or transcription. Token use is also mixed. Median hidden-reasoning tokens fall in both comparisons, while summed use changes by $-16.7\%$ in the exact-output study and $+1.9\%$ in the visible-rationale study.

The pilot identifies a useful distinction for future experiments. Supplying a valid relation and applying it accurately to a concrete input are separate demands. Its small, format-sensitive gains establish neither a reliable accuracy improvement nor an efficiency benefit. A larger study can test the proposed compression mechanism with unchanged-output controls, placebo guidance, more models, and adequately powered case-level comparisons.

\FloatBarrier
\Needspace{22\baselineskip}
\section{Real-Case State Transformations}
\label{app:construction-example}

For the \texttt{abc310\_e} case in Figure~\ref{fig:running-example}, the inside arm increments a loop counter and inserts the following checkpoint before the clean loop body.

\begin{minipage}{\linewidth}
\begin{verbatim}
count += 1
if count == 502:
    emit({"ans": ans, "cnt1": cnt1, "dp1": dp1})
    raise SystemExit
\end{verbatim}
\end{minipage}

The emitted values therefore describe the state after 501 complete iterations. The post arm executes all 1,002 iterations, then inserts the following checkpoint immediately after the loop.

\begin{minipage}{\linewidth}
\begin{verbatim}
if count > 1000:
    emit({"ans": ans, "cnt1": cnt1, "dp1": dp1})
    raise SystemExit
\end{verbatim}
\end{minipage}

The actual benchmark sources use deterministic canonical JSON serialization. The abbreviated \texttt{emit} calls above stand for the exact committed serialization code. The corresponding oracle values are \texttt{\{"ans":63250,"cnt1":500,"dp1":500\}} and \texttt{\{"ans":251502,"cnt1":2,"dp1":2\}}.

\end{document}